\documentstyle[aps,prl,epsf,multicol,rotate,array]{revtex}
\newcommand{\lrule}{ \end{multicols} \noindent
  \rule{0.5\textwidth}{0.1mm}\rule{0.1mm}{3pt}\newline }
\newcommand{\rrule}{ \noindent \parbox{\textwidth}{
  \hfill\rule[-3pt]{0.1mm}{3pt}\rule{0.5\textwidth}{0.1mm}}
  \begin{multicols}{2}\noindent }
\def\UPt{\mbox{UPt$_3$ }}
\def\SR{\mbox{Sr$_2$RuO$_4$ }}
\def\SRend{\mbox{Sr$_2$RuO$_4$}}
\def\AF{{\scriptsize{\mbox{sfl}}}}
\def\bmqc{{\mbox{\boldmath$Q$}}}
\def\bmq{{\mbox{\boldmath$q$}}}
\def\bmp{{\mbox{\boldmath$p$}}}
\def\bmd{{\mbox{\boldmath$d$}}}
\def\bmpp{{\mbox{\boldmath$p$}}^\prime}
\def\bmvf{{\mbox{\boldmath$v$}}_{\!f}}
\def\bmpf{{\mbox{\boldmath$p$}}_{\!f}}
\def\sbmpf{{\mbox{\boldmath$p$}}_{\!f}}
\def\bmpfp{{\mbox{\boldmath$p$}}_{\!f}^\prime}
\def\sbmpfp{{\mbox{\boldmath$p$}}_{\!f}^\prime}

\def\ud{{\psi_{\!\uparrow}\psi_{\!\downarrow}}}
\def\du{{\psi_{\!\downarrow}\psi_{\!\uparrow}}}
\begin{document}
\draft
\date{\today}

\title{Mixed-Parity Superconductivity in \SRend
}

\author{M.~Eschrig$^a$, J.~Ferrer$^b$, and M.~Fogelstr\"om$^c$}
\address{
$^a$Materials Science Division, Argonne National Laboratory, Argonne,
Illinois 60439}
\address{
$^b$Departamento de Fisica, Facultad de Ciencias, Universidad de Oviedo,
E-33007 Oviedo, Spain}
\address{
$^c$Institute for Theoretical Physics, Chalmers University of
Technology and G\"oteborg University, S-41296, G\"oteborg, Sweden
}
\maketitle
\begin{abstract}
We show that in \SR the Fermi surface geometry as inferred from
angle resolved photoemission experiments has important implications
for a pairing interaction dominated
by incommensurate, strongly anisotropic, spin fluctuations.
For a spin fluctuation spectrum consistent with 
inelastic neutron scattering experiments 
the system is close to an accidental degeneracy
between even parity spin singlet and odd parity
spin triplet channels.
This opens the possibility of a
mixed parity order parameter state in \SRend. We determine
the stable and metastable order parameter phases at low temperatures
and discuss especially phases with order parameter nodes.
\end{abstract}

\begin{multicols}{2}
The study of the superconducting material \SR 
\cite{Maeno1994} 
has attracted considerable
experimental and theoretical attention because of its peculiar
low energy electronic properties \cite{Sigrist1999}.
It is a rare example of a multiband
superconductor with possibly spin triplet pairing symmetry\cite{Ishida98}.
Based on muon spin relaxation measurements\cite{Luke98} an early
suggestion for the order parameter symmetry was the 
time reversal symmetry breaking
`$p_x\!\pm\! i p_y$'-state\cite{Sigrist1999}.
Recent experiments \cite{Nishizaki1999,Suderow1998,Bonalde2000}
on the contrary favor order parameters with nodes \cite{Nodes_2000}.
The spin excitation spectrum in \SR as studied by
inelastic neutron scattering (INS) reveals spin fluctuations
in a crossover regime between
ferromagnetic and anti-ferromagnetic fluctuations.
The dominant contributions to the spin susceptibility, $\chi(\bmq)$, are
located around $\bmqc=(1 \pm \delta ,  1 \pm \delta ,0)\; \pi/a$, 
with an incommensurability $\delta \approx 0.4 $\cite{Sidis1999}.
Similar results have also been suggested theoretically finding 
$\delta \approx 1/3$\cite{Mazin1999}. 
It was suggested that in the cross-over regime between
ferromagnetic and anti-ferromagnetic spin fluctuations the order parameter
changes from $p$-wave to $d$-wave\cite{Mazin1999,Kuwabara2000}.
As experimental results become more reliable, it becomes possible to test 
if a pairing mechanism via incommensurate
spin-fluctuations is consistent with
a) the spin fluctuation spectrum as measured in INS,
b) the electronic structure as tested by
angle resolved photoemission and de Haas-van Alphen spectroscopy,
and c)
the experimental implications for the order parameter symmetry.

In this Letter we study the possible pairing symmetries for the order
parameter in
\SR resulting from a pairing interaction dominated by incommensurate,
strongly anisotropic, spin fluctuations.
Given the measured Fermi surface geometry
in \SR \cite{ARPES,dHvA}, we find 
that the system is close
to two accidental degeneracies between the triplet $E_{2u}$ superconducting state
and one of the singlet superconducting states either in
the $B_{1g}$ (predominantly $d$-wave) channel or in the
$A_{1g}$ (predominantly $g$-wave) channel.
This opens the possibility of a superconducting state with mixed 
parity order parameter in \SRend.
The possibility of such superconducting states was
discussed some time ago in connection with \UPt\cite{Garg1993}.
We also study the different possible stable and metastable 
low-temperature phases
in \SR for the three cases that either of the
three relevant order parameter symmetries $E_{2u}$, $B_{1g}$, or
$A_{1g}$ is dominant. 
We find both nodeless solutions and order parameters with nodes.
A dominant triplet 
order parameter is
supported only in a surprisingly small incommensurability region 
near the experimentally determined value $\delta \approx 0.4$.

NMR experiments suggest a strong anisotropic spin susceptibility,
with only the $\chi_{zz}$ peaked around the incommensurate wave vector
$\bmqc$\cite{Mukuda1998}.
Thus, we consider 
an anisotropic model with $\chi^z\equiv \chi_{zz} \gg
\chi_{xx}=\chi_{yy} \equiv \chi^{\perp}$ and 
neglect off-diagonal components 
($\chi_{xy}=0$). 
For $\chi^z$ we assume the following form 
($a=1$ in our units)
\begin{equation}
\chi^z(\bmq)=\sum_{\delta_{x,y}=\pm \delta}
\frac{\chi_Q/4}{1+4\xi^2_{\AF}(\cos^2\frac{q_x-\delta_x}{2}+\cos^2\frac{q_y-\delta_y}{2})} \, .
\label{susceptability}
\end{equation}
This model susceptibility has three parameters.
The overall magnitude, $\chi_Q$, determines the coupling strength
in the dominant pairing channel, and thus the superconducting
transition temperature $T_{c0}$.
The other two parameters, the spin-spin correlation
length, $\xi_{\AF}$, and the degree of incommensuration, $\delta$,
determine the relative coupling strengths in the different symmetry
channels,
defining the symmetry of dominant and sub-dominant components.
The specific form of the susceptibility 
allows a smooth cross-over from
anti-ferromagnetic fluctuations, $\bmqc_{AF}=(1,1,0) \; \pi/a$,
to ferromagnetic,
$\bmqc_{FM}=(0,0,0)$, by tuning $\delta$ from $0$ to $1$.
Extracting the 
values of $\xi_{\AF}$ and $\delta$ from the INS data \cite{Sidis1999}
gives
$\xi_{\AF}\approx4.0 a$ and $\delta\approx0.4 $.
The effective pairing interaction via spin fluctuation exchange
is determined by the coupling function
$g(\bmp ) \chi^{i} (\bmp -\bmpp ) g(\bmpp ) $.
The coupling between spin fluctuations and
quasiparticles, $g(\bmp )$, is approximated in what follows by a constant, $g$.
To reproduce the structure of the experimentally
probed  three-sheet Fermi surface\cite{ARPES,dHvA} 
we use the tight-binding dispersions
\begin{equation}
\epsilon^{i}_{\bmp}=2 t^i_x \cos p_x+2 t^i_y \cos p_y -4
t^{\prime,i}\cos p_x \cos p_y -\mu^i.
\label{dispersion}
\end{equation}
The band index $i$ labels the $xy$, $xz$ and $yz$ bands. The
parameters of the dispersions
$(t^i_x, t^i_y, t^{\prime,i}, \mu^i)$  are
taken from Refs. \cite{Mazin1997}.
Additionally, a hybridization of the $xz$ and the $yz$ bands is
given by $t_{\perp}=0.1$ eV\cite{Morr2000}.

The order parameter in weak coupling BCS theory is determined by the well-known 
non-linear BCS self-consistency equation. 
The smallness of the gap in \SR of about 1 meV \cite{Laube2000} allows us
to restrict momentum summations in this gap equation
to the vicinity of the Fermi surface.
In a standard way this procedure leads to a replacement of
the momentum sum by a Fermi surface average, $\langle \cdots
\rangle_{\sbmpf }$,
weighted with the angle resolved density of states,
$N(\bmpf) = N_f n(\bmpf)=N_f\,|\bmvf(\bmpf )|^{-1}/
\mbox{$\langle |\bmvf(\bmpfp )|^{-1} \rangle_{\sbmpfp }$}$.
$N_f$ is the total density of states at the Fermi level
and $\bmvf $ is the Fermi velocity. 
We obtain $N_f=1.78$ states/eV per spin and unit cell.
Introducing the dimensionless coupling function
$\bar{\chi}=N_f \; g^2\chi $, 
the weak coupling gap equation reads
\begin{eqnarray}
\Delta_{\alpha \beta} (\bmpf)&&= -\,
\sum_{i=\{x,y,z\}}
\sum_{\gamma \delta }
\nonumber \\
&& \times \Big\langle
\sigma_{\alpha \gamma }^i \bar{\chi}^i(\bmpf-\bmpfp) \sigma_{\beta\delta }^i
\; n(\bmpfp )
f_{\gamma \delta }(\bmpfp )
\Big\rangle_{\sbmpfp }
\label{gapxx}
\end{eqnarray}
where $f_{\gamma \delta }(\bmpf)=
T\sum_{\epsilon_{n}}^{\epsilon_{c}} \int d\xi_{p}
F_{\gamma \delta }(\bmpf, \epsilon_{n};\xi_{p})$, $F$ is the anomalous
propagator,
$\epsilon_n$ fermionic Matsubara frequencies, and
$\epsilon_{c}$ is the usual frequency cut-off.
The anisotropic interaction in Eq. (\ref{gapxx}) breaks spin
rotational symmetry, but
since each $\bmpf$-state is doubly Kramers-degenerate in zero-field, we
can still
decompose $\Delta_{\alpha \beta}$ and $F_{\gamma \delta }$ into
(pseudo-)spin singlet ($s$)
and (pseudo-)spin triplet ($t$)
components\cite{Sauls1994} and arrive, for $x=s,t$, at
\begin{equation}
\Delta_{x} (\bmpf)=
-\Big\langle
V_x(\bmpf-\bmpfp) \; n(\bmpfp ) \; f_x(\bmpfp )
\Big\rangle_{\sbmpfp }
\label{gapx}
\end{equation}
with $V_s=\bar{\chi}^{\perp}+\bar{\chi}^z/2$,
$V_{t}^{z}=-\bar{\chi}^{\perp}+
\bar{\chi}^z/2$, $V_{t}^{\perp}=-\bar{\chi}^z/2$.
Isotropic spin fluctuations ($\chi^{\perp}=\chi^z$) support triplet
superconductivity only for nearly ferromagnetic enhancement.
In the case of extreme anisotropy the coupling functions for
singlet and triplet pairing with $\bmd$-vector in
$\hat{z}$-direction have equal sign and magnitude,
$V_s=V_t^z=\bar{\chi}^z/2$. 
In addition, a second triplet-pairing
state with $\bmd\perp\hat{z}$ is possible.
It couples via $V_t^{\perp}=-\bar{\chi}^z/2$.

In order to study the symmetry of the superconducting state near $T_{c0}$
we determine numerically the eigenvalue spectrum of the integral kernel in Eq. \ref{gapx},
following Ref. \cite{Buchholtz1995}.
The resulting complete orthogonal set of basis functions 
${\cal Y}^\Gamma_\mu(\bmpf)$ can be classified according to the 
irreducible representations $(\Gamma)$ of the crystal group
$\mbox{D}_{4h}$. The corresponding eigenvalues, $\lambda^\Gamma_\mu $,
determine the coupling constants for the $\mu $th basis function 
in the symmetry channel $(\Gamma )$\cite{Yip1993}.
The most attractive (negative) eigenvalue in representation $(\Gamma )$, 
$\lambda^\Gamma= \min_\mu (\lambda^\Gamma_\mu )$
is eliminated in favor of a transition
temperature for order parameter symmetry $(\Gamma )$ in the usual way,
$T_{c,\Gamma}=1.13 \epsilon_{c}\exp (-1/|\lambda^\Gamma |)$.

The dominant coupling constant $\lambda^{\Gamma_0} $ 
determines the superconducting transition
temperature $T_{c0}$ and the symmetry $(\Gamma_0 )$ of the superconducting
phase near $T_{c0}$.
Once
\begin{figure}
\centerline{
\rotate[r]{ \epsfxsize=0.28\textwidth{\epsfbox{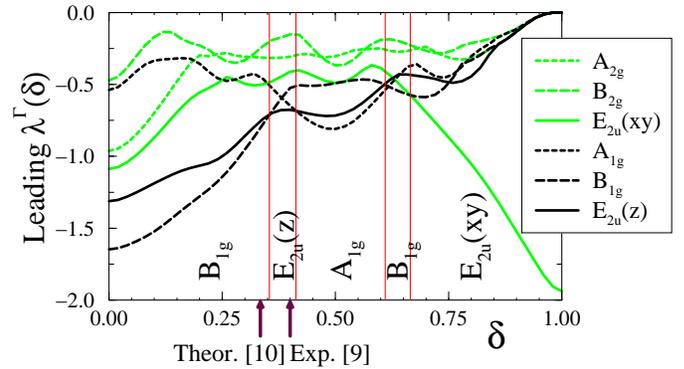}}}
}
\caption[]{
The most attractive eigenvalues, $\lambda^{\Gamma }$, for the different irreducible
$D_{4h}$ crystal symmetry group representations, $(\Gamma )$,
as a function of the
magnetic incommensuration parameter $\delta$ at $\xi_{\AF}=4a$. 
Even-parity representations are $A_{1g}$, $A_{2g}$, $B_{1g}$, and $B_{2g}$.
The odd-parity states have either
$\bmd\parallel\hat{z}$ ($E_{2u}(z)$) or $\bmd\perp\hat{z}$  ($E_{2u}(xy)$).
The regions, separated by vertical lines, 
are marked by the representation of highest $T_{c}$.
}
\label{Eigenvalues}
\end{figure}
\noindent
an order parameter component in a certain representation $(\Gamma )$ nucleates,
the other components $\mu (\Gamma )$ in the same
representation are usually induced by the presence of the first component.
Thus, physical transitions between different superconducting phases
only occur when additional symmetries are spontaneously broken.
Such subdominant transitions are suppressed below the value $T_{c,\Gamma}$
in the presence of a dominant order parameter. 
In the following we study which of the different possible phases nucleates first at
given $\xi_{\AF}$ and $\delta$ and determine points of accidental degeneracy
between two different order parameter phases as a function of these parameters.
In Fig. \ref{Eigenvalues} we show 
the dependence of the attractive eigenvalues on $\delta$ for $\chi^\perp=0$, fixing
$\xi_{\AF}$ at $4.0 a$, a value which should closely correspond to the
actual value in \SRend \cite{Sidis1999}.
The relevant phases near $T_{c0}$ are the 
even-parity one-dimensional
irreducible representations $A_{1g}$ and $B_{1g}$,
which give spin-singlet superconductivity, 
and the odd-parity two-dimensional irreducible
representation, $E_{2u}$, rendering a spin-triplet channel for superconductivity.
At small values of incommensuration, up to $\delta\approx 0.35 $, the
system prefers the $B_{1g}$-pairing channel.
Above this point there is a region,
$0.35  \lesssim \delta \lesssim 0.42  $, of spin-triplet
pairing in the $E_{2u}$ channel with $\bmd\parallel \hat{z}$.
Beyond $\delta \approx 0.42$, the pairing state is spin-singlet with
$A_{1g}$ symmetry,
bounded by a narrow region of $B_{1g}$-pairing 
starting around $\delta\approx0.61$.
In the large-$\delta$ range there is again triplet pairing, but now with $\bmd\perp \hat{z}$.
The two 
accidental degeneracy points of interest for us are $B_{1g}\oplus E_{2u}$ for
an incommensuration $\delta \approx  0.35 $
and $ A_{1g}\oplus E_{2u}$ for $\delta \approx  0.42 $.
Both points are remarkably close to the experimental value of $\delta \approx 0.4 $ and
the theoretically predicted value of $\delta \approx 0.33$.
Calculations with an additional component $\chi^\perp $ resulted into
accidental degeneracies even closer to each other.
We also performed calculations for $\delta \approx 0.4 $ and varying $\xi_{\AF}$ showing
that the presence of the accidental degeneracies near $\delta =0.4 $ is
a robust feature for $\xi_{\AF} > 2a $.

In the following we study the possible low temperature superconducting phases
close to the accidental degeneracies. We concentrate on the three cases
where either of the three symmetry channels is slightly dominant. We chose
$\delta =0.3 $ for a dominant $B_{1g}$ channel, $\delta = 0.4 $ for
a dominant $E_{2u}$ channel, and $\delta =0.45 $ for a dominant
$A_{1g}$ channel.
To determine the superconducting state at low temperatures we solve
the non-linear gap equation (\ref{gapx}).
Expanding the order parameter with respect to the set of basis functions
${\cal Y}^{\Gamma }_\mu(\bmpf)$
we obtain the
order parameter components $\Delta^{\Gamma }_\mu$ for each representation.
In the case of a mixed parity state a mixture
of even parity, $\Delta^{(e)}(\bmpf)$, and odd-parity, $\Delta^{(o)}(\bmpf)$, 
basis functions occurs as 
\begin{equation} 
\Delta_\pm(\bmpf)=
\sum_\mu \Delta^{(e)}_\mu {\cal Y}^{(e)}_\mu(\bmpf)
\pm \sum_\mu \Delta^{(o)}_\mu {\cal Y}^{(o)}_\mu(\bmpf).
\end{equation}
Once parity is broken, each of the doublets, $\ud$ and $\du$, acquire
separate order parameters, $\ud$ with $\Delta_+(\bmpf)$ and $\du$ 
with $\Delta_-(\bmpf)$, and $\Delta(\bmpf)$ has the required
anti-symmetry 
since $\Delta_\pm(\bmpf)=-\Delta_\mp(-\bmpf)$. 
As there may be several possible superconducting states, each state being a
local minima in the free energy,
we compute the free energy of each candidate state using the
Serene-Rainer free energy functional
\cite{Serene1983}, and select the state of lowest energy as the
low-temperature phase.
As the temperature evolution of the order parameter depends on
$T_{c,\Gamma }/T_{c0}=\exp(1/\lambda^{\Gamma }-1/\lambda^{\Gamma_0})$, it is
dependent
on the value of the most negative eigenvalue $\lambda^{\Gamma_0}$, which is determined
by the parameter $g^2\chi_Q$. 
We have chosen $g^2\chi_Q/(2\pi)^2=1$,
leading to eigenvalues $\lambda^\Gamma \lesssim 1$, appropriate for weak coupling.

\begin{figure}
\centerline{
\epsfxsize=0.20\textwidth{\epsfbox{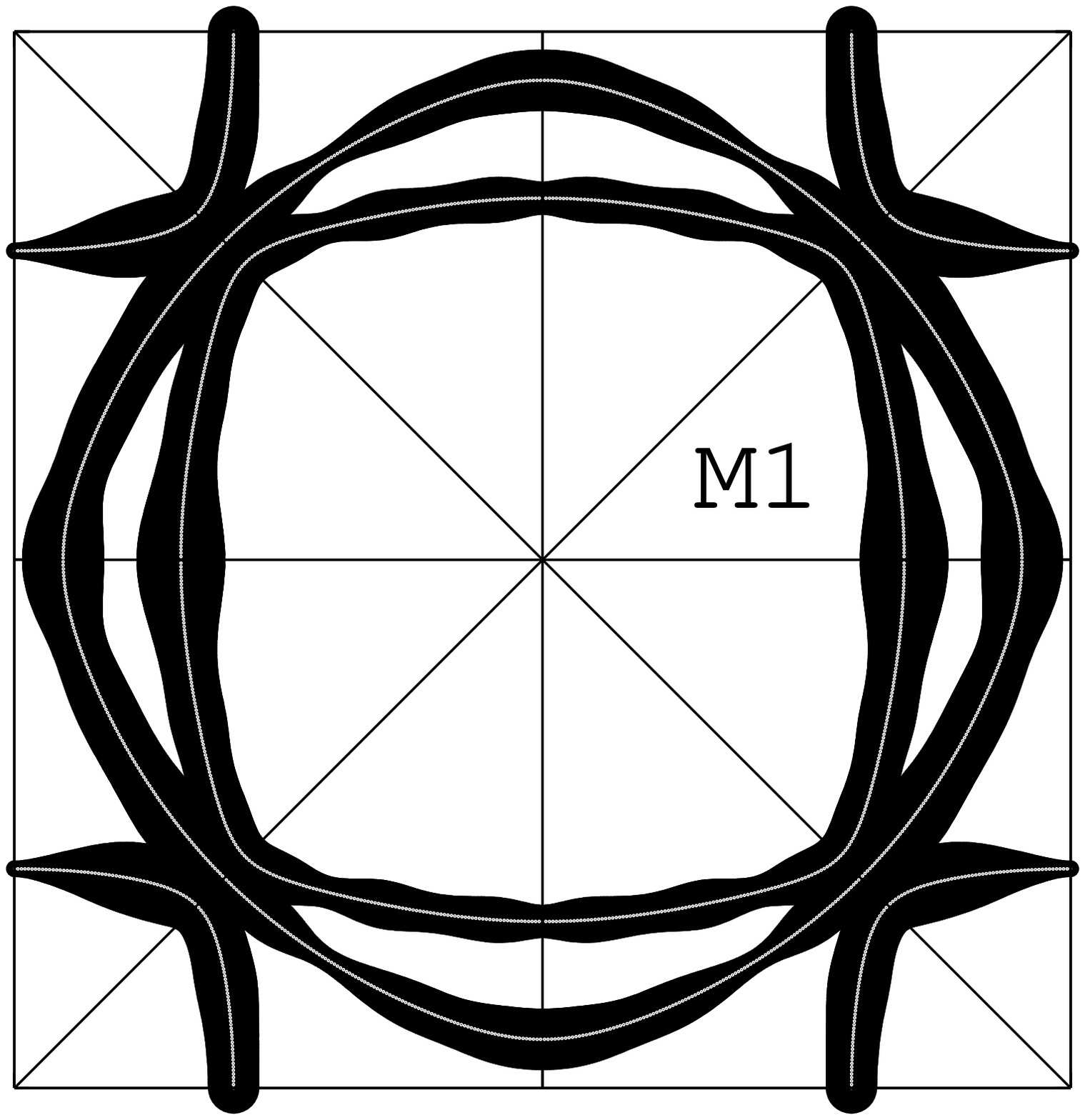}}
\epsfxsize=0.20\textwidth{\epsfbox{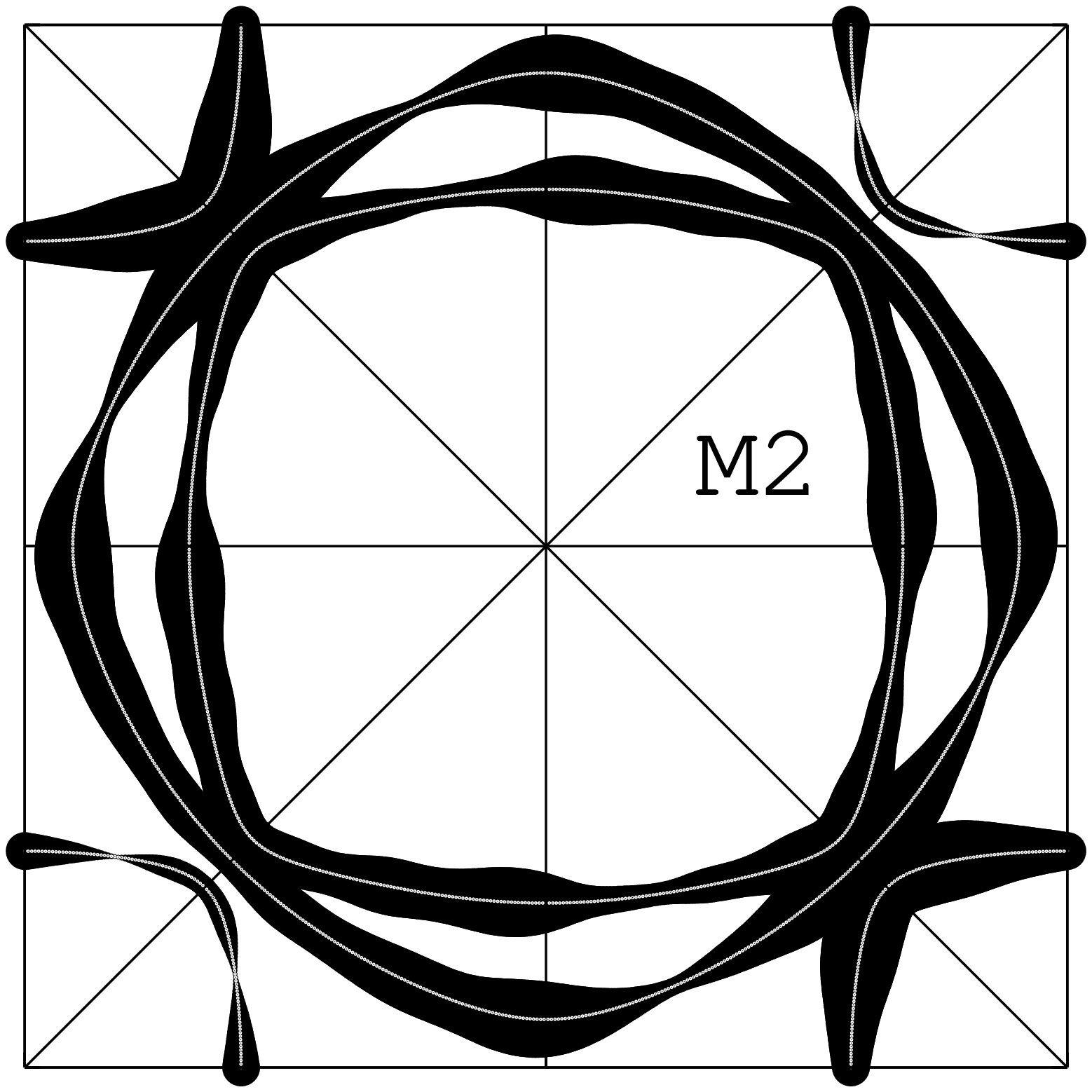}}}
\centerline{
\epsfxsize=0.20\textwidth{\epsfbox{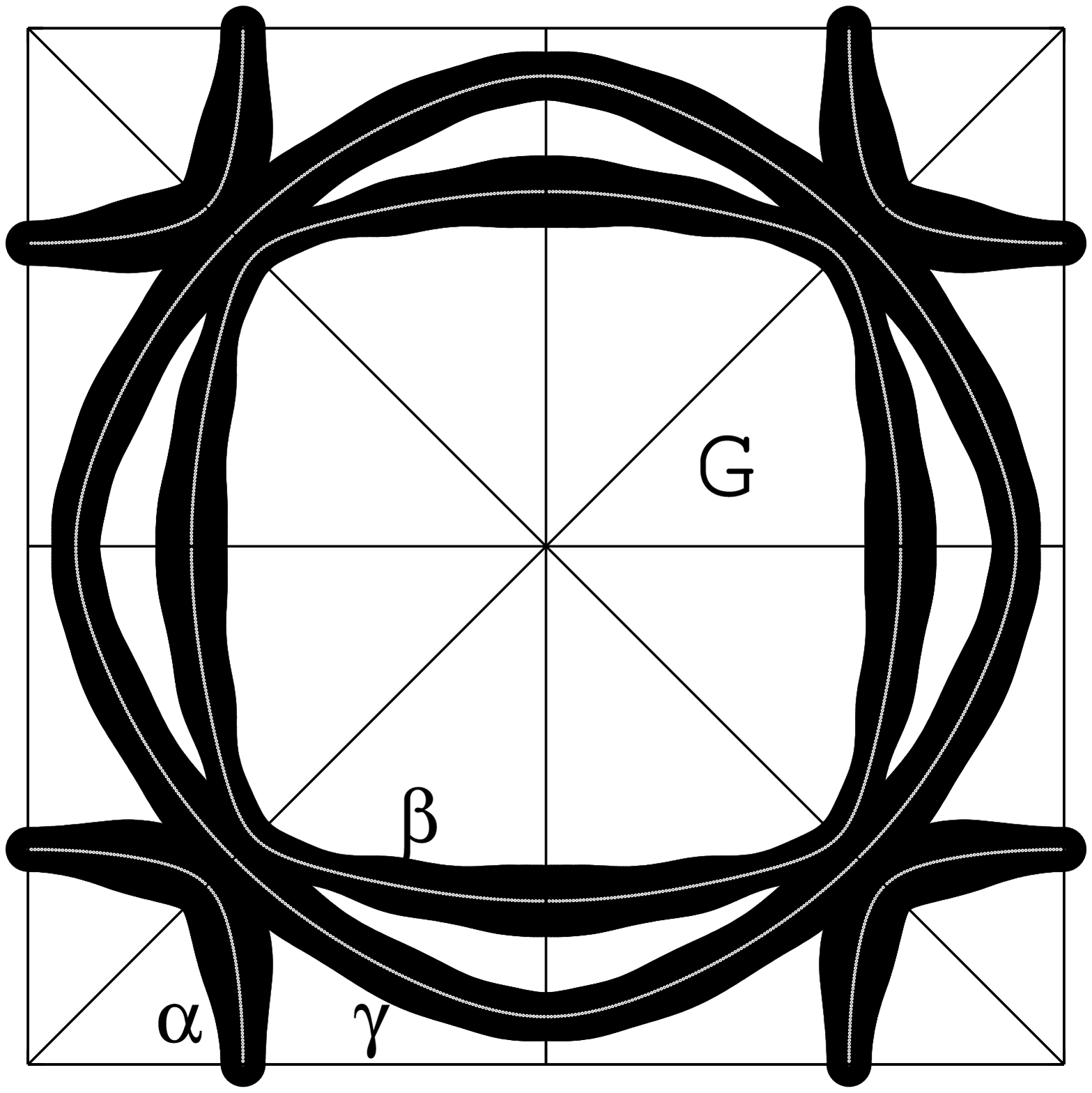}}
\epsfxsize=0.20\textwidth{\epsfbox{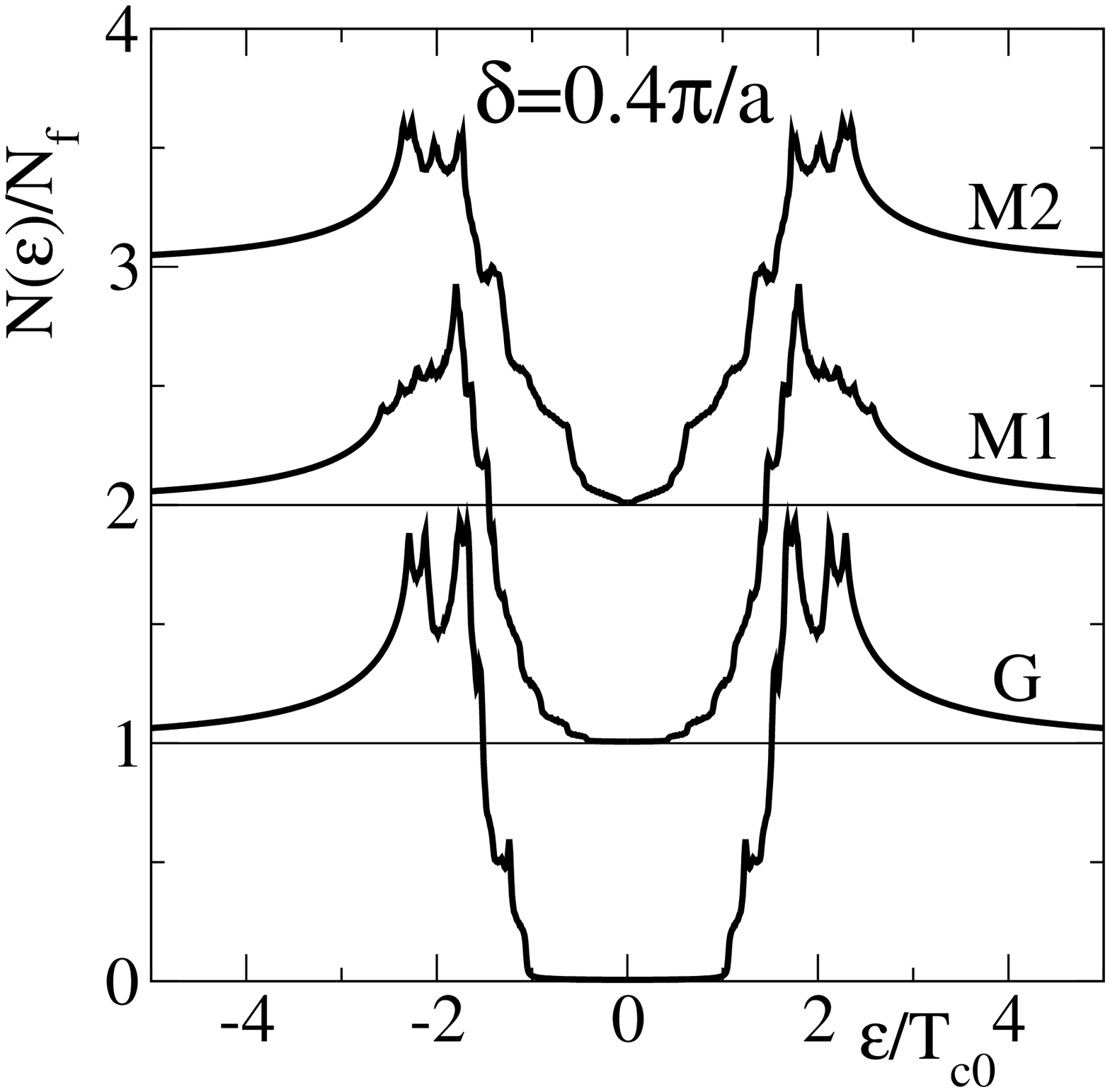}}}
\caption[10]{
The magnitude of the order parameter, $|\Delta_{+}(\bmpf)|$, 
at zero temperature
for the superconducting ground state (G)
and the two metastable states (M1, M2), calculated
for $\delta=0.4$ and $\xi_{\AF}=4a$.
The three Fermi sheets ($\alpha,\beta$ and $\gamma$) 
are marked by thin white lines and $|\Delta_{+}(\bmpf)|$ by the
thickness of the black lines. 
The ground state is of pure $E_{2u}$ symmetry, the metastable states
mix $E_{2u}$, $A_{1g}$, and $B_{1g}$,
resulting in mixed parity states. 
The corresponding total density of states, 
$N(\epsilon )$, for each state is shown in
the lower right panel.
}
\label{sopE}
\end{figure}

In Fig. \ref{sopE} we plot
the magnitude of the superconducting gap as a function of position on the Fermi
surface, 
$|\Delta_{+}(\bmpf)| $, that
minimizes the free energy at $\delta=0.4$ and with
$\xi_{\AF}=4a$. 
We find three local minima: one ground state (G) and
two metastable states (M1 and M2).
The free energy difference at zero temperature
between M1 and G corresponds 
to only 9\% of the ground state condensation energy (it amounts to  22\% for M2).
G is of pure $E_{2u}$-symmetry.  M1
is symmetric around the $p_x$ and $p_y$ axis and has points with
small gap values on the $\alpha $-sheet. M2
is symmetric around the diagonals $p_x\pm p_y$ and has nodes on the
$\alpha $ sheet. Also
shown in Fig. \ref{sopE} is the total (angle-averaged)
density of states (DOS) at the Fermi surface.
The DOS is fully gapped for the ground state. The first metastable  state shows
a DOS with a much smaller excitation gap, 
and the DOS for the second metastable state shows
low energy nodal excitations originating from the $\alpha $ sheet.
The metastable states break spin-rotation symmetry and parity.
The amplitudes $\Delta^{(e,o)}_\mu$ have a zero relative phase 
for basis functions within the same parity, and
a relative phase difference
of $\pm \pi/2$ for basis functions with opposite parity;
thus, 
\noindent
$\Delta_-(\bmpf)=\Delta_+(\bmpf)^{\ast}$.
In the ground state the odd-parity basis functions
with same eigenvalue have a $\pi/2$ relative phase difference,
giving a `$p_x\!\pm\! i p_y$'-state.
We emphasize: all three solutions
break time-reversal symmetry.

\begin{figure}
\centerline{
\epsfxsize=0.20\textwidth{\epsfbox{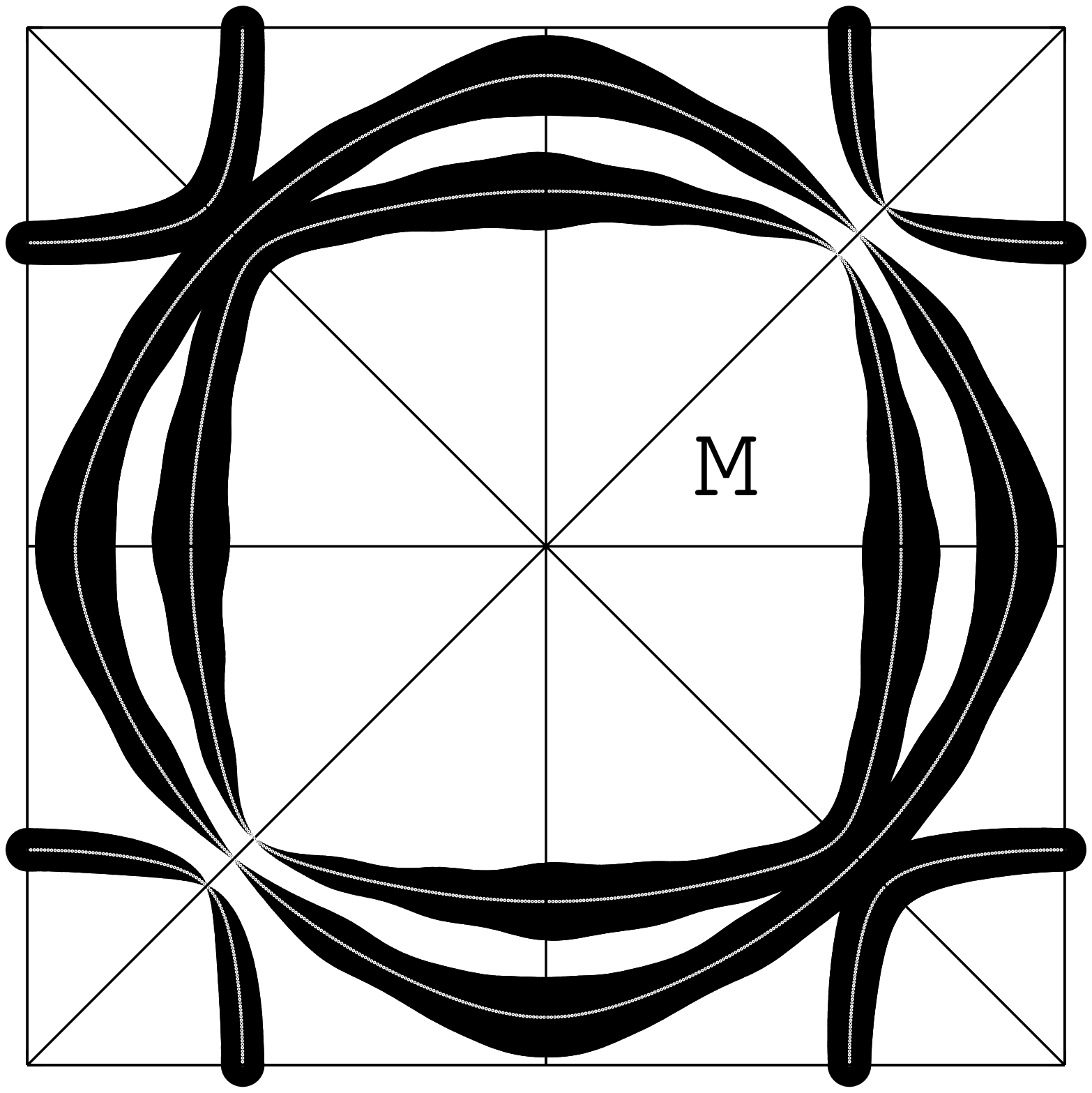}}
\epsfxsize=0.20\textwidth{\epsfbox{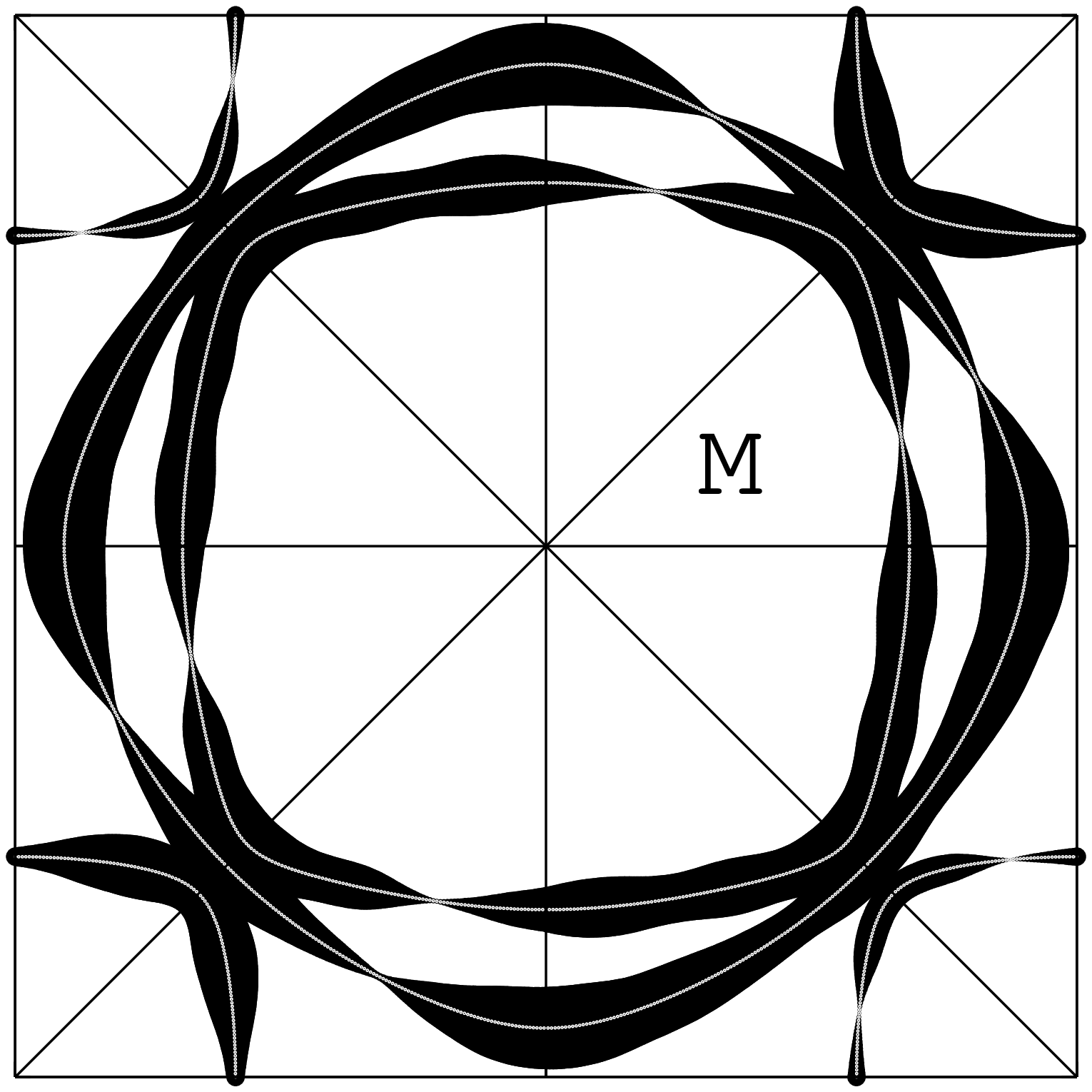}}}
\centerline{
\epsfxsize=0.20\textwidth{\epsfbox{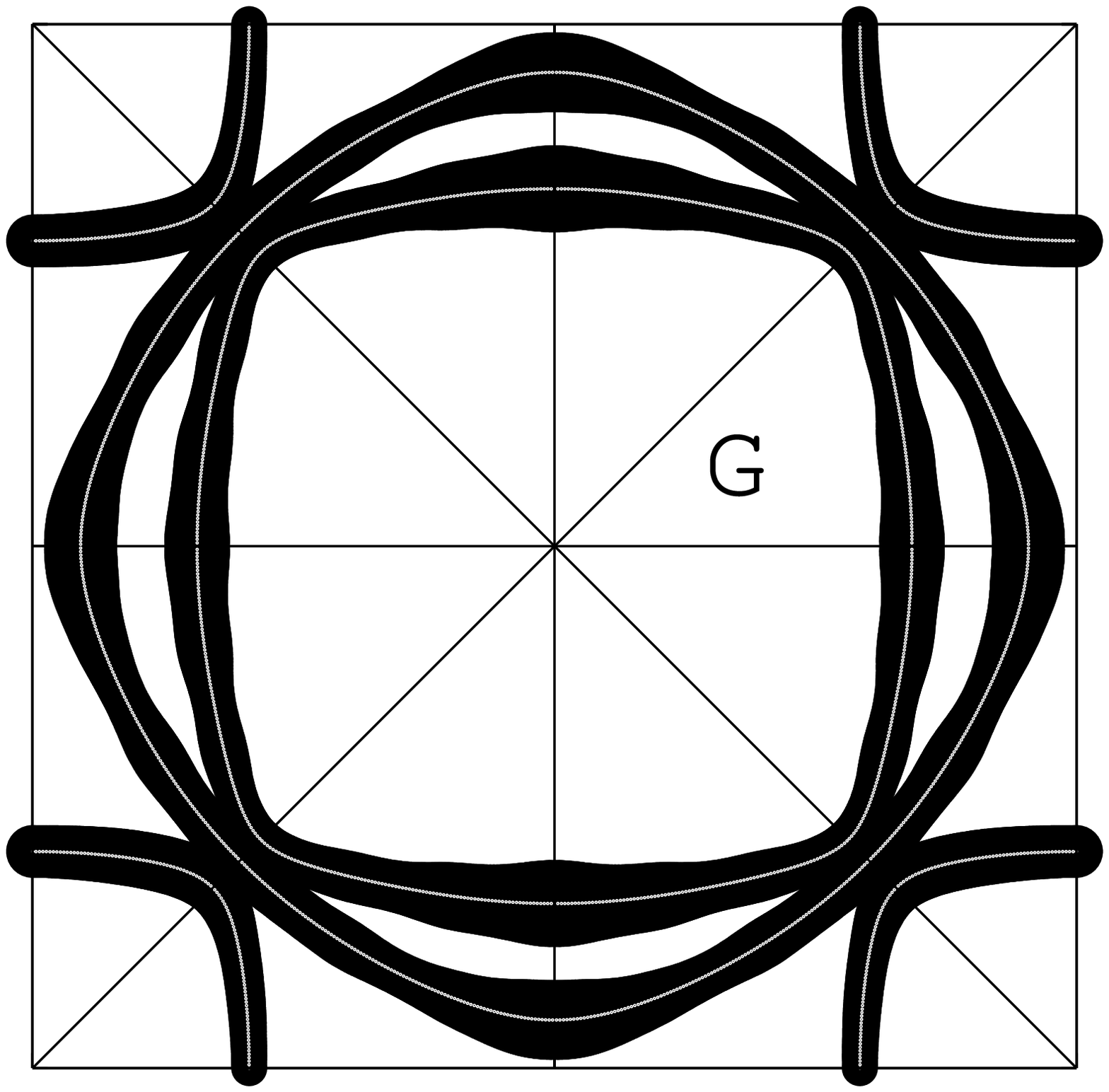}}
\epsfxsize=0.20\textwidth{\epsfbox{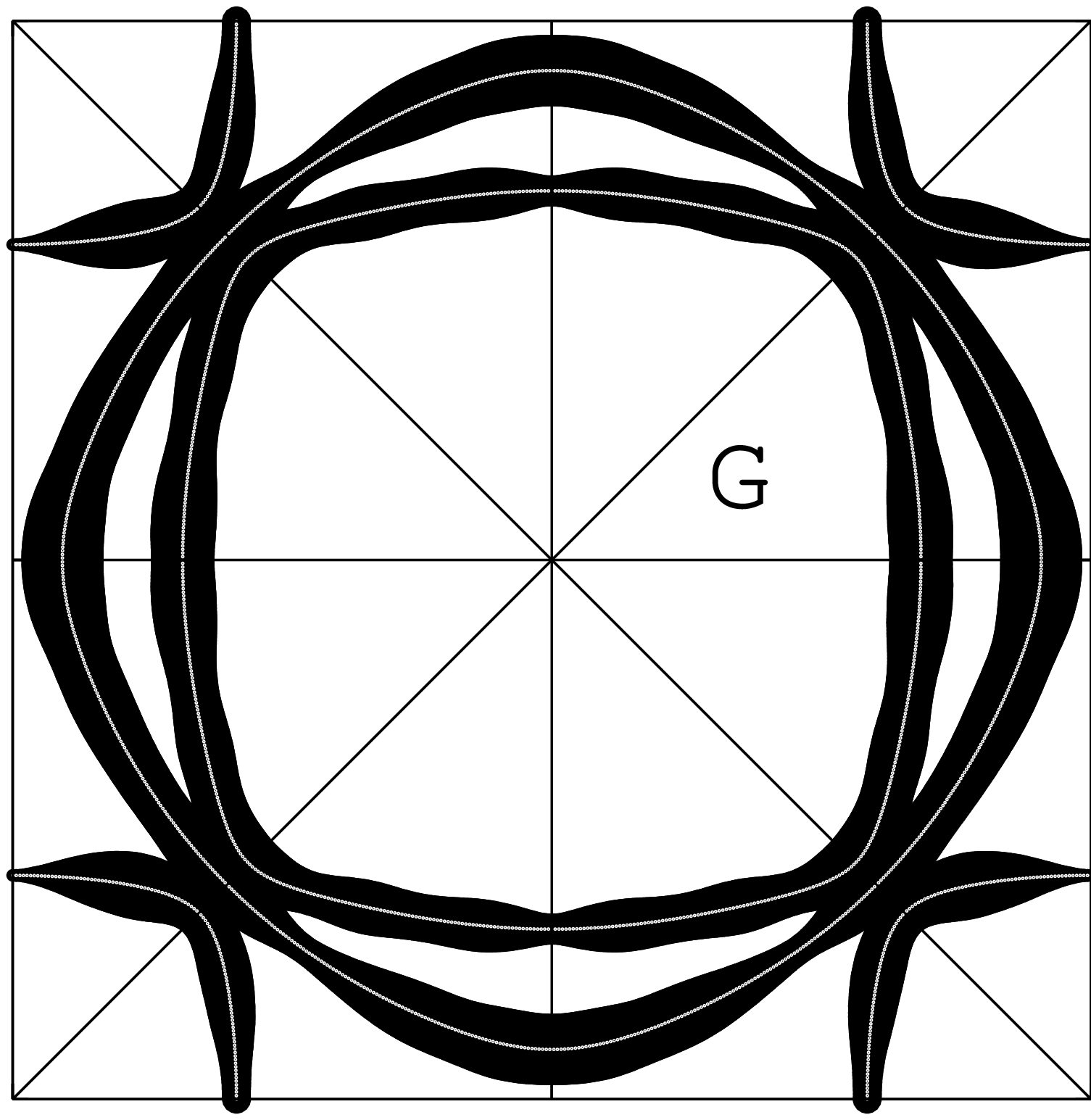}}}
\centerline{
\epsfxsize=0.19\textwidth{\epsfbox{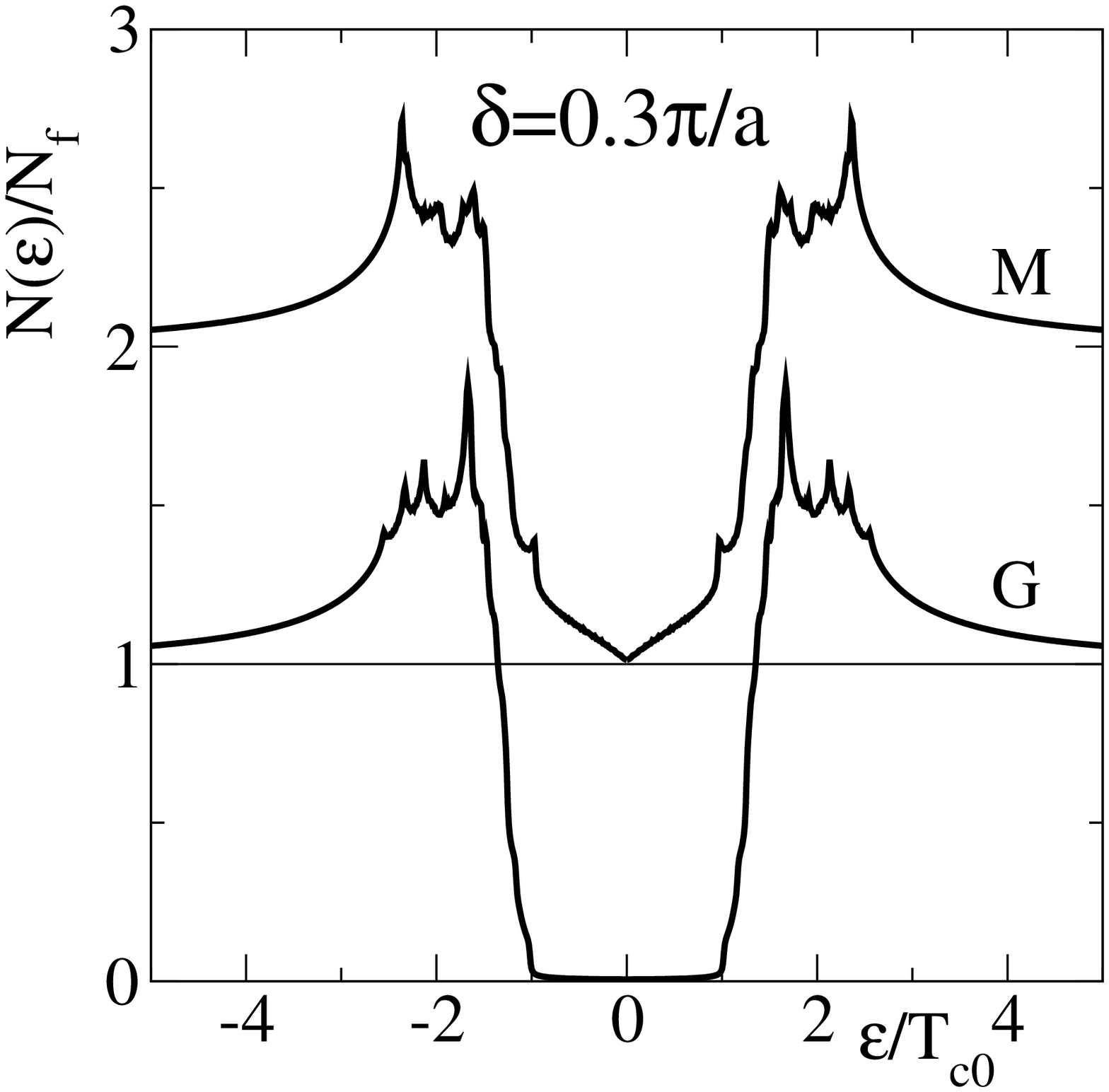}}
\epsfxsize=0.19\textwidth{\epsfbox{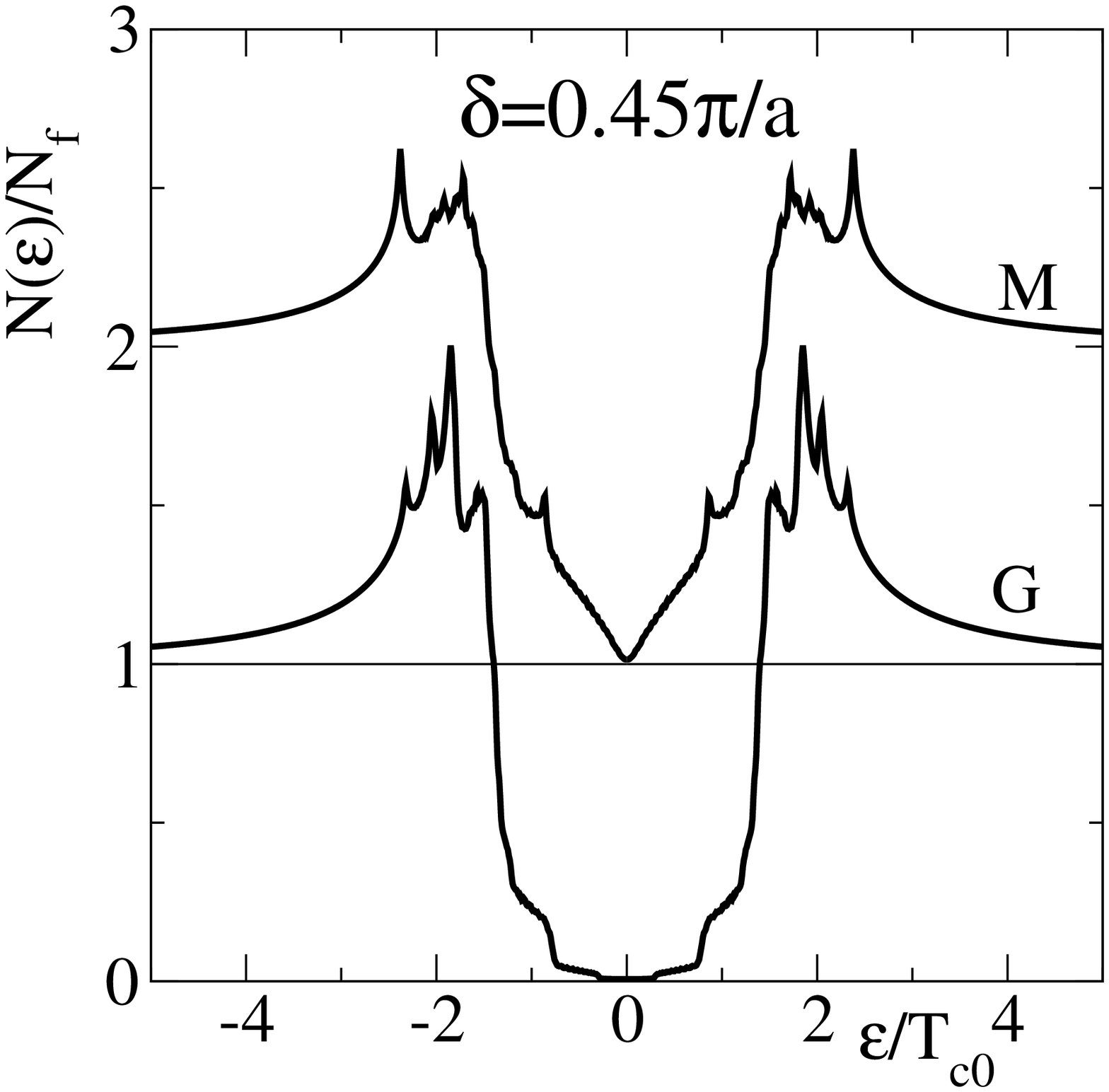}}}
\caption[]{
The same as in Fig. \ref{sopE} for $\delta=0.3$ (left) and
for $\delta=0.45$ (right).
Top to bottom: 
gap magnitude of the metastable state (M), gap magnitude of the ground state (G), 
and the density of states for each corresponding state.
}
\label{sops}
\end{figure}

Going to a $\delta$ where 
superconductivity nucleates first in either the $B_{1g}$ or the $A_{1g}$ channel,
we find that already the ground states are mixed parity states.
The lowest energy state is
a nodeless time reversal symmetry breaking state with relative phases similar to
the M1 state discussed above.
For both $\delta=0.3$ and $\delta=0.45$ we find metastable
states, with nodes on all three Fermi surface sheets, that are of
slightly higher
free energies (9.0\% and 15.6\% respectively) than the nodeless ground states.
In Fig. \ref{sops} we show $|\Delta_+(\bmpf)|$
and the DOS for both states at $\delta=0.3$ and $\delta=0.45$. 
For the metastable states 
the odd-parity components can
develop higher order nodes, and remarkably, even
whole arcs with vanishing or very small order parameter magnitude
at the positions of the nodes in the even-parity components. 
At $\delta =0.3 $, for instance, the odd parity component has nodes
along a diagonal in the Brillouin zone, coinciding with the $d$-wave nodes of the
even parity component. In this case such arcs of almost zero $\Delta^{(o)}(\bmpf )$
occur on the $\gamma $-sheet, extending all over
the two quadrants which contain the nodes of the total order parameter.

The two components of the odd-parity order parameter transform like a vector in momentum
space. Its direction is determined by the anisotropy introduced 
by a) the normal state
DOS and b) the presence of the even-parity 
\noindent
$B_{1g}$ or $A_{1g}$ components. The mixed parity solutions for
each incommensurability correspond to the alignment of this vector with
high symmetry directions in the Brillouin zone. 
Because the energy difference
between the ground state and the state with
nodes is only a fraction of the ground
state condensation energy, fluctuations in the direction of this vector
can be responsible for the presence of nodal excitations.
Analyzing data of
specific heat measurements \cite{Nishizaki1999} and of
thermal conductivity \cite{Suderow1998}, prompts
that the \SR order parameter should
have nodes \cite{Nodes_2000}. This conclusion is
further strengthened by recent measurements of the
penetration depth showing a non-exponential
low-temperature behavior\cite{Bonalde2000}.
Based on the presented calculations, this implies that a
mixed parity superconducting state, breaking
spin-rotation symmetry, may be a candidate state for \SRend, and maybe even
stabilized by additional interactions in the Hamiltonian not considered here.

In conclusion, we have shown that in the parameter region supported by
experimental results even parity singlet and odd parity triplet pairing compete
to have the highest transition temperature.
Given the spin-spin correlation length
of several lattice constants, as suggested by experiment, for
an incommensuration near $\delta =  0.35 $ 
superconductivity nucleates 
in an accidentally degenerate 
$B_{1g}\oplus E_{2u}$ state and near $\delta =  0.42 $
the state is $A_{1g}\oplus E_{2u}$. 
Both states are of mixed parity
and break the $D_{4h}$ crystal symmetry and time reversal symmetry.
Within our model we find as ground states
time reversal symmetry breaking nodeless states.
However, close in free energy there exist
metastable states of the order parameter with nodes, which may be stabilized
by additional interactions.

We would like to thank Juana Moreno, Michael R. Norman, 
Stellan \"Ostlund, and
James A. Sauls
for clarifying discussions.
This work is supported by the
TMR network ``Phase coherent transport of hybrid nanostructures'' contract
No. FMRX-CT96-0042,
the
U.S. Department of Energy, Basic Energy Sciences, Contract No.
W-31-109-ENG-38 (ME),
and by the Swedish Natural Science Research Council (MF).

\end{multicols}
\end{document}